\documentclass{article}  
\usepackage{bigsky2009}
\usepackage{graphicx}
\usepackage{epstopdf}
\def\eq#1{{(\ref{#1})}}

\newcommand{\beq}{\begin{equation}}
\newcommand{\eeq}{\end{equation}}
\newcommand{\ben}{\begin{eqnarray*}}
\newcommand{\een}{\end{eqnarray*}}
\frompage{000} \topage{000}                                              
\def\rightmark{Topologically induced local {\cal P} and {\cal CP} violation in hot QCD}
\def\leftmark{D. E. Kharzeev}
 
\title{Topologically induced local {\cal P} and {\cal CP} violation in hot QCD} 
\authors{
{Dmitri E. Kharzeev$^{1,2}$ %
}\\[2.812mm]
{\normalsize
\hspace*{-8pt}$^1$ Nuclear Theory Group, \\
Physics Department\\
Brookhaven  National Laboratory\\
Upton, New York 11973, USA\\[0.2ex] 
\hspace*{-8pt}$^2$ Physics Department\\ 
Yale University\\
 New Haven, CT 06520-8120, USA
}}
 
\abstract{
Very stringent experimental bounds exist on the amount of P and CP violation in strong interactions. 
Nevertheless, the presence of non-Abelian topological solutions and the axial anomaly make the issue of CP invariance in QCD non-trivial ("the strong CP problem"). Even in the absence of a global P and CP violation the fluctuations of topological charge in the QCD vacuum are expected to play an important role in the breaking of chiral symmetry, and in the mass spectrum and other properties of hadrons. 
Here I argue that topological fluctuations in hot QCD matter can become directly observable in the 
presence of a very intense external magnetic field by inducing {\it local} P- and CP- odd effects. 
These local parity-violating phenomena can be described by using the Maxwell-Chern-Simons, or axion, electrodynamics as an effective theory. Local P and CP violation in hot QCD matter can be observed in experiment through the "chiral magnetic effect" -- the separation of electric charge along the axis of magnetic field that is created by the colliding relativistic ions. There is a recent evidence for the electric charge separation relative to the reaction plane of heavy ion collisions from the STAR Collaboration at RHIC.}
 
\begin{document}
 
\maketitle
\setcounter{page}{1}

\section{Introduction}
The non-Abelian nature of QCD is responsible for chirality non-conservation in this theory. The change of chirality is caused by the topological gluon field configurations that couple to quarks via the axial anomaly. Indeed, the classical Yang-Mills equations (non-Abelian analogs of Maxwell equations) possess non-trivial vacuum solutions -- instantons -- 
corresponding to the mapping 
of the $SU(2)$ subgroup of the gauge group $SU(3)$ onto the group of three-dimensional rotations $S_3$ \cite{Belavin:1975fg}.    
Instantons thus couple rotations in space to rotations in the space of color. In the presence of fermions, this property of instantons combined with the axial anomaly \cite{anomaly} causes non-conservation of chirality that would otherwise be forbidden by the conservation of angular momentum -- but the coupling of angular momentum to color allows to compensate the flip of spin by a rotation in color space.   
 In Minkowski space-time instantons describe the tunneling transitions between the states with different topological Chern-Simons numbers \cite{Chern:1974ft} $\nu$ of the $SU(2) \leftrightarrow S_3$ mapping \cite{'tHooft:1976fv,Jackiw:1976pf,Callan:1976je}; for a review, see \cite{Schafer:1996wv}.

At finite 
 temperature, the transition between the vacuum states with different topological numbers can occur not only through quantum tunneling, but can also be induced by a classical thermal activation process, through a "sphaleron" \cite{Klinkhamer:1984di}. In electroweak theory sphaleron transitions cause the baryon number violation and may be responsible for at least a part of the observed baryon asymmetry in the Universe \cite{Kuzmin:1985mm}; for a review, see \cite{Rubakov:1996vz}. Sphalerons are also expected to play a role in QCD plasma \cite{McLerran:1990de} where they induce the quark chirality  non-conservation. Unlike for the instanton transitions \cite{Gross:1980br}, the rate of the sphaleron transitions $\Gamma$ is not exponentially suppressed at weak coupling $g$, and in Yang-Mills theory with $N$ colors is proportional to \cite{Arnold:1996dy,Huet:1996sh,Bodeker:1998hm} 
\beq\label{weak}
 \Gamma = const \times (g^2 N)^5 \ln(1/g^2 N) \ T^4
\eeq
  (with a numerically large pre-factor \cite{Moore:1997sn}). Sphalerons describe a random walk in the topological number; in a volume $V$ and for a large time period $T$ we get the topological number $\langle \nu^2 \rangle =  \Gamma\ V\ T$. The diffusion of topological charge of course is expected to occur not only at weak coupling; while one cannot yet  compute the corresponding rate analytically in QCD, lattice calculations indicate a large rate at experimentally accessible temperatures of $T = 200 \div 300$ MeV \cite{Moore:1997sn}. The distributions of topological charge measured on the lattice close to the deconfinement phase transition are quite broad (for a review, see \cite{Vicari:2008jw}); this feature of the QCD plasma is consistent with the expectations based on effective theories, and can cause P and CP odd effects in heavy ion collisions \cite{Kharzeev:1998kz,Kharzeev:1999cz}. Topological transitions do not require thermal equilibrium, and can be induced in the pre-equilibrium stage of a heavy ion collision  \cite{Kharzeev:2001ev,Kharzeev:2000ef,Shuryak:2000df,Nowak:2000de,Lappi:2006fp}.
  
  A valuable insight is offered also by the ${\cal N} =4$ maximally super-symmetric Yang-Mills theory where the topological charge diffusion rate can be evaluated in the strong coupling limit through the AdS/CFT correspondence, with the following result \cite{Son:2002sd}: 
 \beq\label{topdiff}
 \Gamma = (g^2 N)^2/(256\pi^3)\ T^4,
 \eeq
which shows that the topological transitions become more frequent at strong coupling, even though the dependence on the coupling is weaker than suggested by \eq{weak}; note that the large $N$ behavior is the same in the weak and strong coupling limits ($\sim N^0$). It is worth mentioning that the calculation \cite{Son:2002sd} of the topological charge diffusion rate \eq{topdiff} is analogous to the evaluation \cite{Policastro:2001yc} of shear viscosity of the strongly coupled ${\cal N} =4$ SUSY Yang-Mills theory. The only difference is that on the super-gravity side in the latter case one is dealing with the propagation of gravitons (excited by the components of the energy-momentum tensor on Minkowski boundary) in $AdS_5$ space, and in the former -- of the axions excited by the operator of the topological charge. 

Topological fluctuations are believed to play an important role in the structure of QCD vacuum and in the properties of hadrons (for a review, see \cite{Schafer:1996wv}). They also open the possibility of  P and CP violation in QCD ("the strong CP problem")\footnote{This possibility is probably not realized in the present-day Universe -- the experimental bounds from the electric dipole moments on the amount of P and CP violation in QCD  are very stringent.}. However until now all of the evidence for the topological effects in QCD from experiment, however convincing, has been indirect. Here I will present the arguments \cite{Kharzeev:2004ey,Kharzeev:2007tn,Kharzeev:2007jp,Fukushima:2008xe} for the possibility to observe the topological effects in QCD directly in the presence of very intense external electromagnetic fields. In particular, the coupling of topological gluon field configurations to electromagnetism induced by the axial anomaly leads to the separation of electric charges in the presence of a strong external magnetic field ("the chiral magnetic effect"). The electromagnetic fields of the required  strength can be created in heavy ion collisions; there is a recent evidence 
for the charge separation effect from STAR experiment at RHIC \cite{Voloshin:2008jx}.

\section{Topologically induced effects in ${\rm\bf QCD} \times {\rm\bf QED}$: \\ Maxwell-Chern-Simons theory}
 
\subsection{The Lagrangian} 
 
Consider QCD coupled to electromagnetism; the resulting theory possesses $SU(3) \times U(1)$ gauge symmetry:
$$
{\cal L}_{\rm QCD+QED} =  -{1 \over 4} G^{\mu\nu}_{\alpha}G_{\alpha \mu\nu}  + \sum_f \bar{\psi}_f \left[ i \gamma^{\mu} 
(\partial_{\mu} - i g A_{\alpha \mu} t_{\alpha} -  i q_f A_{\mu}) -  m_f \right] 
\psi_f  - 
$$
\beq\label{qcd+qed}
- {\theta \over 32 \pi^2}  g^2 G^{\mu\nu}_{\alpha} \tilde{G}_{\alpha \mu\nu} - \frac{1}{4}F^{\mu\nu}F_{\mu\nu},
\eeq 
where $A_{\mu}$ ($A_{\alpha \mu}$) and $F_{\mu\nu}$ ($G_{\mu\nu\alpha}$) are the electromagnetic (color) vector potential and the corresponding field strength tensor, and $q_f$ are the electric charges of the quarks.   

Let us discuss the electromagnetic sector of the theory  \eq{qcd+qed} at large distances. Electromagnetic fields will couple to the electromagnetic currents $J_\mu = \sum_f  q_f \bar{\psi}_f \gamma_\mu \psi_f$.  
In addition, the $\theta$-term in \eq{qcd+qed} through the quark loop will induce  the coupling of $F \tilde{F}$ to the QCD topological charge. We will introduce an effective pseudo-scalar field $\theta = \theta(\vec x, t)$ (playing the role of the axion field) and write down the resulting effective Lagrangian as
\beq\label{MCS}
{\cal L}_{\rm MCS} = - \frac{1}{4}F^{\mu\nu}F_{\mu\nu} - A_\mu J^\mu - \frac{c}{4}\ \theta \tilde{F^{\mu\nu}}F_{\mu\nu},
\eeq
where 
\beq\label{coef}
c = \sum_f q_f^2 e^2 / (2\pi^2). 
\eeq

This is the Lagrangian of Maxwell-Chern-Simons, or axion, electrodynamics in $(3+1)$ dimensions that has been discussed previously in \cite{Wilczek:1987mv,Carroll:1989vb,Sikivie:1984yz}. 
If $\theta$ is a constant, then the last term in \eq{MCS} represents a full divergence 
\beq\label{an_ab}
\tilde{F^{\mu\nu}} F_{\mu\nu} = \partial_\mu J_{CS}^\mu
\eeq
of the Chern-Simons current 
\beq\label{topdiv1}
J_{CS}^{\mu} = \epsilon^{\mu\nu\rho\sigma} A_{\nu} F_{\rho\sigma}.
\eeq
Being a full divergence, this term 
does not affect the equations of motion.

The situation is different if the field $\theta = \theta(\vec x, t)$ varies in space-time.      
Indeed, in this case we have
\beq
\theta \tilde{F^{\mu\nu}} F_{\mu\nu} = \theta \partial_\mu J_{CS}^\mu = \partial_\mu\left[\theta J_{CS}^{\mu}\right] - \partial_\mu \theta  J_{CS}^{\mu}.
\eeq
The first term on r.h.s. is again a full derivative and can be omitted; introducing notation
\beq
P_\mu = \partial_\mu \theta = ( \dot{\theta}, \vec P )
\eeq
we can re-write the Lagrangian \eq{MCS} in the following form:
\beq\label{CS}
 {\cal L}_{\rm MCS} = - \frac{1}{4}F^{\mu\nu}F_{\mu\nu} - A_\mu J^\mu + \frac{c}{4} \ P_\mu J^\mu_{CS}.
\eeq
Since $\theta$ is a pseudo-scalar field, $P_\mu$ is a pseudo-vector; as is clear from   \eq{CS}, 
it plays a role of the potential coupling to the Chern-Simons current \eq{topdiv1}. However, unlike the vector potential $A_\mu$, $P_\mu$ is not a dynamical variable and is a pseudo-vector that is fixed by the dynamics of chiral charge -- in our case, determined by the fluctuations of topological charge in QCD.
\vskip0.3cm

\subsection{Maxwell-Chern-Simons equations}   
   
Let us write down the Euler-Lagrange equations of motion that follow from the Lagrangian \eq{CS},\eq{topdiv1}  
(Maxwell-Chern-Simons equations):
\beq
\partial_\mu F^{\mu\nu} = J^\nu - P_\mu \tilde{F}^{\mu\nu}.
\eeq
The first pair of Maxwell equations (which is a consequence of the fact that the fields are expressed through the vector potential) is not modified:
\beq
\partial_\mu \tilde{F}^{\mu\nu} = J^\nu.
\eeq   
It is convenient to write down these equations also in terms of the electric $\vec E$ and magnetic $\vec B$ fields:
\beq\label{MCS1}
\vec{\nabla}\times \vec{B} - \frac{\partial \vec{E}}{\partial t} = \vec J + c \left(\dot{\theta} \vec{B} - \vec{P} \times \vec{E}\right), 
\eeq
\beq\label{MCS2}
\vec{\nabla}\cdot \vec{E} = \rho + c \vec{P} \cdot \vec{B},
\eeq
\beq\label{MCS3}
\vec{\nabla}\times \vec{E} +  \frac{\partial \vec{B}}{\partial t} = 0,
\eeq
\beq\label{MCS4}
\vec{\nabla}\cdot \vec{B} = 0,
\eeq
where $(\rho, \vec J)$ are the electric charge and current densities.
One can see that the presence of Chern-Simons term leads to essential modifications of the Maxwell theory. 

\subsection{The Witten effect}

As a first application, let us consider a magnetic monopole  in the presence of finite $\theta$ angle. It was shown by Witten \cite{Witten:1979ey} that at finite $\theta$ angle the monopole acquires electric charge $e \theta / 2 \pi$. We will now derive this result following Wilczek \cite{Wilczek:1987mv} by considering the equations of axion electrodynamics. In the core of the monopole $\theta=0$, and away from the monopole $\theta$ acquires a finite non-zero value -- therefore within a finite domain wall we have a non-zero $\vec P = \vec{\nabla} \theta$ pointing radially outwards from the monopole. According to \eq{MCS2}, the domain wall thus acquires a non-zero charge density $c  \vec{\nabla} \theta \cdot \vec{B}$. An integral along $\vec P$ (across the domain wall) yields $\int dl\ \partial \theta / \partial l = \theta$, and the integral over all directions of $\vec P$ yields the total magnetic flux $\Phi$. By Gauss theorem, the flux is equal to the magnetic charge of the monopole $g$, and the total electric charge of the configuration is equal to 
\beq
q = c\ \theta\ g = \frac{e^2}{2\pi^2}\ \theta\ g = \frac{e}{2\pi^2}\ \theta\ (e g) = e\ \frac{\theta}{2 \pi},
\eeq
where we have used an explicit expression \eq{coef} for the coupling constant $c$, as well as the Dirac condition $g e = 4 \pi \times {\rm integer}$.

\section{Chiral magnetic effect}

\subsection{Charge separation}

\begin{figure}[htb]
\noindent
\vspace{-2cm}
\includegraphics[width=14cm]{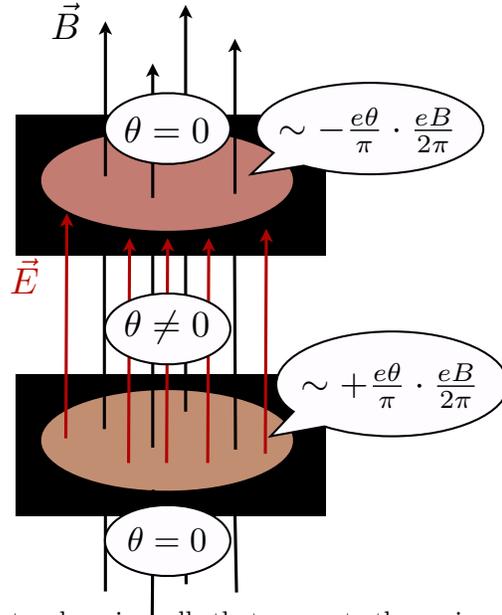}

\vspace{0.5cm}
\caption{
Charge separation effect -- domain walls that separate the region of $\theta \neq 0$ from the outside vacuum with $\theta = 0$ become charged in the presence of an external magnetic field, with the surface charge density $\sim e \theta/\pi \cdot eB/2\pi$.
This induces an electric dipole moment signaling ${\cal P}$ and ${\cal CP}$ violation. 
 }
\label{chargesep}
\end{figure}

Consider now a configuration shown on Fig.\ref{chargesep} where an external magnetic field $\vec B$ pierces a domain with $\theta \neq 0$ inside;  outside $\theta=0$. Let us assume first that the field $\theta$ is static, $\dot\theta = 0$. Assuming that the field $\vec B$ is perpendicular to the domain wall, 
we find from \eq{MCS2} that the upper domain wall acquires the charge density per unit area $S$ of  \cite{Kharzeev:2007tn}
\beq
\left(\frac{Q}{S}\right)_{up}  = +\ c\ \theta B
\eeq
while the lower domain wall acquires the same in magnitude but opposite in sign charge density
\beq
\left(\frac{Q}{S}\right)_{down}  = -\ c\ \theta B
\eeq 
Assuming that the domain walls are thin compared to the distance $L$ between them, we find that 
the system possesses an electric dipole moment
\beq\label{eldip}
d_e = c\ \theta\ (B \cdot S)\ L = \sum_f q_f^2 \ \left(e\ \frac{\theta}{\pi}\right)\ \left(\frac{eB\cdot S}{2\pi}\right)\ L;
\eeq
in what follows we will for the brevity of notations put $\sum_f q_f^2 = 1$; it is easy to restore this factor in front of $e^2$ when needed.
 
Static electric dipole moment is a signature of ${\cal P}$, ${\cal T}$ and ${\cal CP}$ violation (we assume that $\cal{CPT}$ invariance holds). The spatial separation of charge will induce the corresponding electric field $\vec E = c\ \theta\ \vec B$. The mixing of pseudo-vector magnetic field $\vec B$ and the vector electric field $\vec E$ signals violation of ${\cal P}$, ${\cal T}$ and ${\cal CP}$ invariances.

The formula \eq{eldip} allows a simple interpretation: since $eB/2\pi$ is the transverse density of Landau levels of charged fermions in magnetic field $B$, the floor of the quantity $eB\cdot S/2\pi$ (i.e. the largest integer that is smaller than $eB\cdot S/2\pi$) is an integer number of fermions localized on the domain wall. Each fermion species contributes independently to this number as reflected by the factor $N_f$. Again we see that the electric dipole moment \eq{eldip} arises from the electric 
charge $q \sim e \theta/\pi$ that is induced on the domain walls due to the gradient of the pseudo-scalar field $\theta$.  

If the domain is due to the fluctuation 
of topological charge in QCD vacuum, its size is on the order of QCD scale, $L \sim \Lambda_{\rm QCD}^{-1}$, $S \sim \Lambda_{\rm QCD}^{-2}$. This means that to observe an electric dipole moment
in experiment we need an extremely strong magnetic field $eB \sim   \Lambda_{\rm QCD}^{2}$. Fortunately, such fields exist during the early moments of a relativistic heavy ion collision \cite{Kharzeev:2007jp}. Here we have assumed that the domain is static; this approximation requires the characteristic time of topological charge fluctuation $\tau \sim 1/\dot{\theta}$ be large on the time scale at which the magnetic field $B$ varies. This assumption is only marginally satisfied in heavy ion collisions, and so we now need to consider also the case of $\dot\theta \neq 0$.
  
\subsection{The chiral induction}
 
\begin{figure}[htb]
\noindent
\vspace{-1cm}
\includegraphics[width=14cm]{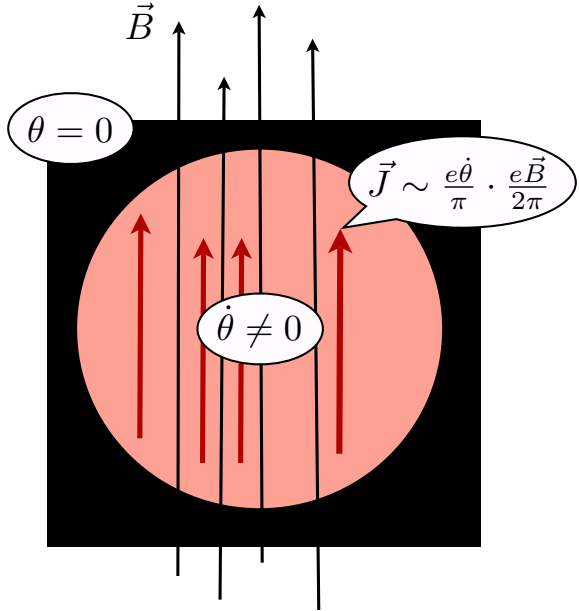}

\vspace{0.4cm}
\caption{
The chiral magnetic effect -- inside a domain with $\dot\theta \neq 0$ an external magnetic field induces an electric current 
$\vec J \sim e \dot\theta/\pi \cdot e \vec B/2\pi$. $\dot \theta \neq 0$ indicates the change of the chiral charge inducing an asymmetry between the left-- and right-- handed fermions inside the domain. Note that the current $\vec J \sim \vec B$ is absent in Maxwell electrodynamics. 
 }
\label{fig:chimagef}
\end{figure}

 Consider now the domain where $| \vec{P} | \ll \dot{\theta}$, i.e. the spatial dependence of $\theta(t, \vec x)$ is much slower than the dependence on time \cite{Kharzeev:2007jp}, see Fig. \ref{fig:chimagef}. Again, we will expose the domain to an external magnetic field $\vec B$ with $\vec{\nabla}\times \vec{B} = 0$, and assume that no external electric field is present.  In this case we immediately get from \eq{MCS1} that there is an induced current \cite{Fukushima:2008xe}
 \beq\label{chimag}
 \vec{J} = - c\ M\ \vec{B} = - \frac{e^2}{2 \pi^2}\ \dot\theta \vec{B}.
 \eeq
Note that this current directed along the magnetic field $\vec B$ represents a ${\cal P}-$, ${\cal T}-$ and ${\cal CP}-$ phenomenon and of course is absent in the "ordinary" Maxwell equations. Integrating the current density over time (assuming that the field $\vec B$ is static) we find that when $\theta$ changes from zero to some $\theta \neq 0$, this results in a separation of charge and the electric dipole moment \eq{eldip}.

 \section{Summary and discussion}

In this talk I argued that the fluctuations of topological charge in the presence of strong magnetic field should give rise to the separation of electric charge along the axis of magnetic field. In heavy ion collisions, the magnetic fields of required strength are produced naturally by the electrically charged ions in the initial state, spectators in the final state, and due to the electric charge asymmetry in the distributions of the produced hadrons \cite{Kharzeev:2007jp}.  The produced magnetic fields are oriented perpendicular to the reaction plane 
(along the system's angular momentum); therefore the effect under discussion should result in the separation of the electric charge with respect to the reaction plane \cite{Kharzeev:2004ey}. Since there is no {\it global} violation of {\cal P} and {\cal CP} invariances in QCD, the sign of the charge asymmetry should fluctuate from event to event. 

The experimental variable sensitive to this effect has been proposed by Voloshin \cite{Voloshin:2004vk}, and the first preliminary results have been 
reported in \cite{Selyuzhenkov:2005xa}. Recently STAR Collaboration has refined and extended this analysis \cite{Voloshin:2008jx}. Numerous mundane backgrounds have been examined, and none of them could so far explain the observed effect \cite{Voloshin:2008jx}. It is clear that a dedicated experimental program of studying topological effects in hot QCD matter is necessary to understand fully this intriguing observation. 

The proposed mechanism requires a sufficiently large energy density for the sphaleron transitions to turn on, and for the quarks to separate by a distance comparable to the system size -- therefore, there has to be a deconfined phase. In addition the system has to be in a chirally symmetric phase -- indeed, in a chirally broken phase, the chiralities of quarks could flip easily causing dissipation of the induced current.  The experimental and theoretical studies of parity--odd charge asymmetries in heavy ion collisions can significantly improve our understanding of the topological structure of QCD, and help to detect the creation of a deconfined and chirally symmetric phase of QCD matter.

  \section*{Acknowledgments}
I am grateful to K. Fukushima, Yu. Kovchegov, A. Krasnitz, E. Levin, L. McLerran, R. Pisarski, M. Tytgat, R. Venugopalan, H. Warringa, and A. Zhitnitsky for enjoyable collaborations on the topics related to this talk. 
The work of D.K. was supported by the U.S. Department of Energy under Contract No. DE-AC02-98CH10886. 

\vfill\eject
\end{document}